%% file: 000-main.tex
\documentclass[sigconf]{acmart}
\input{math_commands.tex}

\usepackage[ruled,vlined]{algorithm2e}
\usepackage{pdfpages}
\usepackage{import}
\usepackage{paralist}
\usepackage{subcaption}
\usepackage{caption}
\usepackage{multirow}
\usepackage{makecell}
\usepackage{threeparttable}

\AtBeginDocument{%
  \providecommand\BibTeX{{%
    \normalfont B\kern-0.5em{\scshape i\kern-0.25em b}\kern-0.8em\TeX}}}

\setcopyright{acmcopyright}
\copyrightyear{2018}
\acmYear{2018}
\acmDOI{10.1145/1122445.1122456}

\acmConference[XXX]{XXXX}{June 03--05, 2021}{XXXX}
\acmBooktitle{WXXXX,
  June 03--05, 2021, XXX, XX}
\acmPrice{15.00}
\acmISBN{978-1-4503-XXXX-X/18/06}

\graphicspath{./img}

\begin{document}


\title{Graph Learning Augmented Heterogeneous Graph Neural Network for Social Recommendation}

\author{Yiming Zhang}
\authornote{Both authors contributed equally to this research.}
\affiliation{%
  \institution{Tongji University}
  \city{Shanghai}
  \country{China}}
\email{2030796@tongji.edu.cn}

\author{Lingfei Wu}
\authornotemark[1]
\affiliation{%
  \institution{JD Silicon Valley Research Center}
  \country{United States}}
\email{lwu@email.wm.edu}

\author{Qi Shen}
\affiliation{%
  \institution{Tongji University}
  \city{Shanghai}
  \country{China}}
\email{1653282@tongji.edu.cn}

\author{Yitong Pang}
\affiliation{%
  \institution{Tongji University}
  \city{Shanghai}
  \country{China}}
\email{1930796@tongji.edu.cn}

\author{Zhihua Wei}
\authornote{Corresponding author.}
\affiliation{%
  \institution{Tongji University}
  \city{Shanghai}
  \country{China}}
\email{zhihua_wei@tongji.edu.cn}

\author{Fangli Xu}
\affiliation{%
  \institution{Squirrel AI Learning}
  \country{United States}}
\email{fxu02@email.wm.edu}

\author{Ethan Chang}
\affiliation{%
  \institution{Middlesex School}
  \country{United States}}
\email{echang@mxschool.edu}

\author{Bo Long}
\affiliation{
  \institution{JD.COM}
  \country{United States}}
\email{bo.long@jd.com}

\renewcommand{\shortauthors}{Zhang, et al.}

\input{00-abs}




\maketitle

\input{0-intro}

\input{1-methods}

\input{2-exp}

\input{3-conclusion}

\bibliographystyle{ACM-Reference-Format}
\bibliography{ref}



\end{document}

%% file: math_commands.tex
\usepackage{amsmath,amsfonts,bm}

\def\eqref#1{equation~\ref{#1}}

\def\1{\bm{1}}

\DeclareMathAlphabet{\mathsfit}{\encodingdefault}{\sfdefault}{m}{sl}
\SetMathAlphabet{\mathsfit}{bold}{\encodingdefault}{\sfdefault}{bx}{n}



%% file: 00-abs.tex
\begin{abstract}
Social recommendation based on social network has achieved great success in improving the performance of recommendation system. Since social network (user-user relations) and user-item interactions are both naturally represented as graph-structured data, Graph Neural Networks (GNNs) have thus been widely applied for social recommendation. 
Despite the superior performance of existing GNNs-based methods, there are still several severe limitations:
(i) Few existing GNNs-based methods have considered a single heterogeneous global graph which takes into account user-user relations, user-item interactions and item-item similarities simultaneously. That may lead to a lack of complex semantic information and rich topological information when encoding users and items based on GNN.
(ii) Furthermore, previous methods tend to overlook the reliability of the original user-user relations which may be noisy and incomplete. 
(iii) More importantly, the item-item connections established by a few existing methods merely using initial rating attributes or extra attributes (such as category) of items, may be inaccurate or sub-optimal with respect to social recommendation.
In order to address these issues, we propose an end-to-end heterogeneous global graph learning framework, namely \emph{Graph Learning Augmented Heterogeneous Graph Neural Network} (GL-HGNN) for social recommendation. GL-HGNN aims to learn a heterogeneous global graph that makes full use of user-user relations, user-item interactions and item-item similarities in a unified perspective.
To this end, we design a Graph Learner (GL) method to learn and optimize user-user and item-item connections separately. 
Moreover, we employ a Heterogeneous Graph Neural Network (HGNN) to capture the high-order complex semantic relations from our learned heterogeneous global graph. 
To scale up the computation of graph learning, we further present the Anchor-based Graph Learner (AGL) to reduce computational complexity. 
Extensive experiments on four real-world datasets demonstrate the effectiveness of our model.
\end{abstract}

%% file: 0-intro.tex
\section{Introduction}
Recent years have witnessed the rapid development of social recommendation, which leverages social network as side information to effectively alleviate the problem of data sparsity \cite{ma2011recommender,wang2019social}. Conceptually, users' preferences are often largely influenced by people around them \cite{fan2019deep,chen2020social}, including parents, friends, classmates, and so on. Therefore, a social recommendation system based on users' social relationships usually significantly improves the quality of recommendations.

Recently, there are a surge of interests in graph neural networks (GNNs) \cite{kipf2016semi,velivckovic2017graph, 2017graphsage,wang2019heterogeneous,xu2020global,pang2021heterogeneous}, which  have been proven to effectively learn node representations from graph-structured data. Since social network (user-user relations) and user-item interactions are both naturally represented as graph-structured data \cite{fan2019graph,wu2020diffnet++}, GNNs have thus been widely employed to learn the  representations of users and items, which has been shown to improve the performance of the recommendation system \cite{zhang2020distilling,wu2019neural,liu2020heterogeneous,he2020game}. Furthermore, in order to enrich the types of the potential graphs and extract richer side information, a few existing works have considered the construction of extra item-item graph structure  \cite{wu2019dual,fan2020graph,huang2021knowledge}.

Despite the promising results current methods have achieved, there are still several severe limitations in their approaches. 
\emph{First}, few existing GNNs-based methods have considered a single heterogeneous global graph which takes into account user-user relations, user-item interactions and item-item similarities simultaneously. As a result, these methods may fail to capture high-order cross-semantic information and limit the delivery of messages.
\emph{Second}, previous methods tend to overlook the reliability of the original user-user social graph that may be noisy and incomplete, partially because that the original connections often only record the social relationships between users but rather reflect the similarity of users preferences. For example: Bob is Ketty’s husband, Jim is Ketty’s colleague, Bob and Jim both like sports, while Ketty likes to read. However, in the user-user graph, Bob and Jim are not directly connected, while they are connected to Ketty respectively. We can learn from this example that there may be conflicts in interests between nearby neighbors, while distinct neighbors could have similar preferences. Such a topology will make the users preferences extracted from user-user graph deviate from the real situation, which may lead to the sub-optimal performance of downstream task.
\emph{Third}, previous methods do not fully exploit the relationships between items. Though a few  existing methods attempt to construct the item-item graph, they only utilize the items' initial rating attributes or extra attributes (such as category) in an ad-hoc fashion, which barely reflects the optimized item-item graph structure with respect to downstream social recommendation. 
\begin{figure*}
    \centering
    \includegraphics[width=0.8\linewidth]{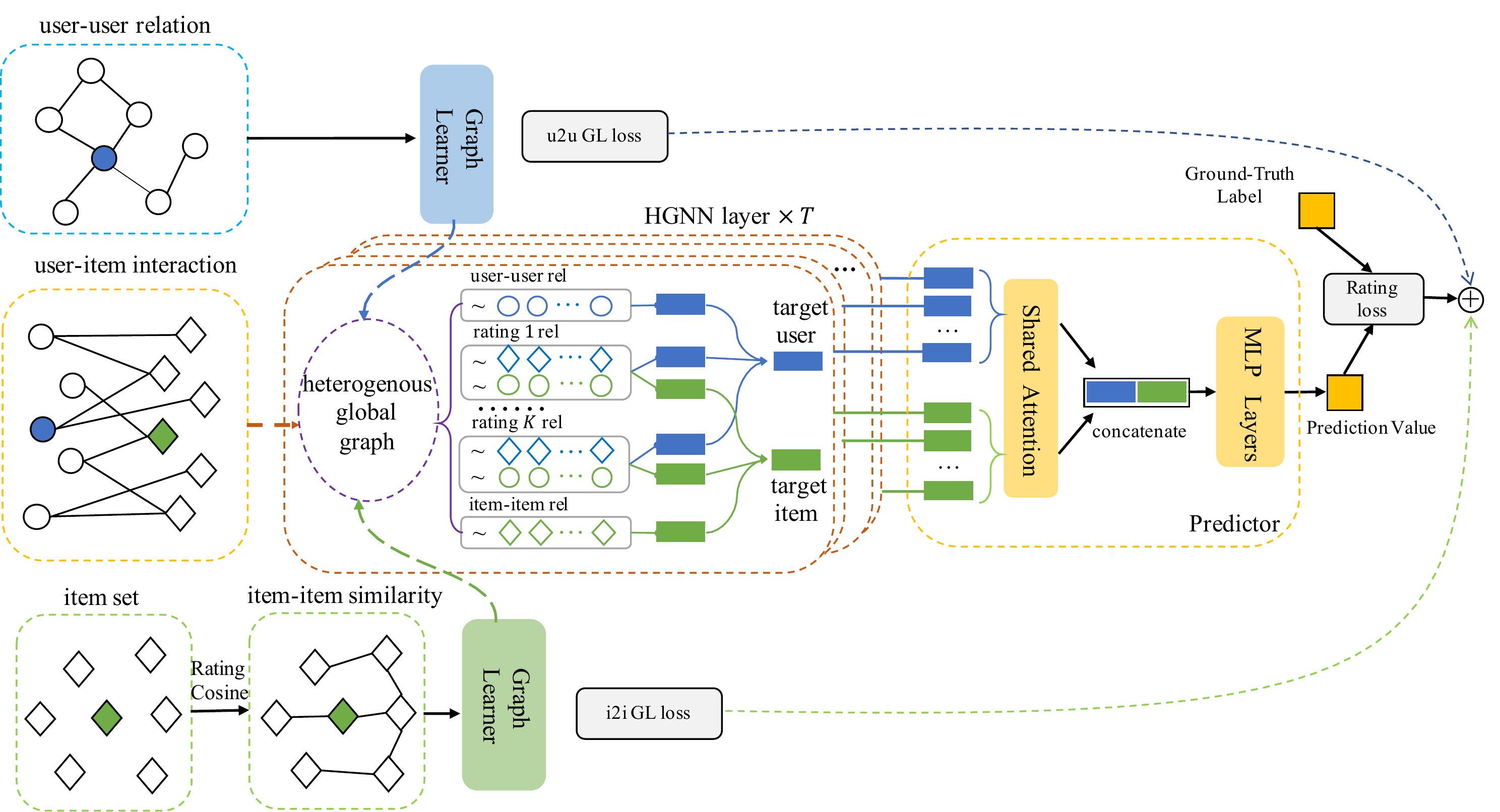}\label{overview}

    \caption{The overview of our model GL-HGNN. We first establish the item-item subgraph structure. We employ the Graph Learner to update and optimize the graph structure. We utilize HGNN to model the global graph to extract complex cross-semantic information. The output embeddings are sent to the predictor for prediction. We design a hybrid loss including Graph Learner loss and rating loss for training.}
    \label{fig:overview}
    \vspace{-0.5cm}
\end{figure*}

In order to address these issues, we propose an end-to-end heterogeneous global graph learning framework, namely \emph{Graph Learning Augmented Heterogeneous Graph Neural Network} (GL-HGNN) for social recommendation. Our GL-HGNN aims to learn a heterogeneous global graph that makes full use of user-user relations, user-item interactions and item-item similarities in a unified perspective.

In order to obtain and optimize heterogeneous global graph structure, we present a Global Graph Learning module. To this end, we first establish item-item subgraph by calculating the similarity of the rating vectors. Then, our proposed Graph Learner (GL) method is employed to extract richer implicit relationships and filter out the explicit noisy edges in user-user relation subgraph and item-item similarity subgraph. Specifically, our GL method can learn the implicit connections between nodes by measuring the embeddings similarity between target nodes in each mini-batch and all nodes. We then combine the learned implicit graph with the initial graph to obtain the refined heterogeneous global with respect to downstream task. 

To capture the high-order complex semantic relations from different types of edges in the heterogeneous global graph, we also present a Heterogeneous Graph Neural Network (HGNN), to model the refined heterogeneous global graph.
Technically, it is crucial to scale up the computation of user-user subgraph and item-item subgraph learning, especially when the number of users or items is very large. To this end, we further utilize anchor-based approximation technique \cite{chen2020iterative} to design a scalable Graph Learner, namely Anchor-based Graph Learner (AGL). By selecting the anchor node set instead of all nodes to calculate the similarity to the target nodes, we can significantly reduce computational complexity. 
In addition, we design a joint learning method and a hybrid loss which considers both graph learner loss and rating loss. Through multiple epochs of optimization, we can get more refined heterogeneous global graph structures with respect to social recommendation, and more reliable vector representations of users and items.
To summarize, we highlight our main contributions as follows:
\begin{itemize}
    \item We construct a heterogeneous global graph with different semantic meta-paths for social recommendation. We propose a novel framework named GL-HGNN to learn the heterogeneous global graph of different relationships in a unified perspective, which can capture the complex semantic relations and rich topological information.
    \item We propose the Global Graph Learning module to construct item-item connections and optimize both user-user and item-item subgraph structures, so as to obtain the refined global graph with respect to the downstream social recommendation. In addition, we design an Anchor-based Graph Learner (AGL) method to scale up the proposed method, which can significantly reduce computational complexity.
    \item We conduct experiments on four real-world datasets. The experimental results demonstrate the effectiveness of our proposed model over state-of-the-art methods, and also verify that our scalable AGL module can reduce the computational costs.
\end{itemize}
\section{Related Works}
\subsection{Social Recommendation} With the popularity of social platforms, social recommendation
has become one of the hottest areas in recommendation research. Early research mainly employed matrix factorization-based methods for recommendation, such as SoRec \citep{ma2008sorec}, TrustFM \cite{10.5555/2540128.2540524} and TrustSVD \cite{guo2015trustsvd}. Recently, deep learning-based methods have become the most successful methods in recommendation research. Plenty of recent works \cite{fan2019deep,shen2020peia,pan2020learning} have applied deep learning to social recommendation tasks and achieved promising performance.

In recent years, a lot of  works \cite{berg2017graph,wu2019neural,fan2019graph} transform user-user relations and user-item interactions to graph-structured data, and employ the graph neural network (GNN) to learn better user and item representations. In addition, to capture connections among items  and 
enhance the performance of social recommendation, several efforts adopted the item attributes to construct item-item graph. For example, GraphRec+ \cite{fan2020graph} and DANSER \cite{wu2019dual} leveraged the item's collaborative information to build item-item graph, while KCGN \cite{huang2021knowledge} utilized inter-dependent knowledge of items to construct graph.

However, few of these methods extract high-order cross-semantic information by modeling a joint heterogeneous global graph, which includes three kinds of meta-paths: user-user, user-item and item-item.
\subsection{Graph Structure Learning}
As GNNs rely on the good quality of the original graph, graph structure learning method was proposed to alleviate this limitation. LDS \cite{franceschi2019learning} proposed to model each edge inside the adjacency matrix. IDGL \cite{chen2020iterative} jointly and iteratively learned graph structure and graph embedding based on node features. HGSL \cite{zhao2021heterogeneous} generated three kinds of graph structures  to fuse an optimal heterogeneous graph.

However, most of these models are applied to node classification or graph-level
prediction tasks. To our best knowledge, we are the first to adopt the graph structure learning to improve the quality of the heterogeneous global graph 
in social recommendation.

%% file: 1-methods.tex
\section{DEFINITION AND PRELIMINARY}

In this paper, we define $U=\{u_1,u_2,...,u_N\}$  and $V=\{v_1,v_2,...,v_M\}$  as the sets of users and items, separately. The user-user relations can be defined as $G_{uu}=\{U,\mathcal{E}_u\}$, in which $\mathcal{E}_u$ is the set of edges, and $(u_{i},u_{n},r_u)$ in $\mathcal{E}_u$ represents $u_{i}$ is related to $u_{n}$. And the user-item interactions can be represented as the user-item graph with $K$ kinds of edges $G_{uv}=\{U,V,\mathcal{E}_r\}$. The edge in $\mathcal{E}_r$ is defined as $(u_i,v_j,r_k)$, which indicates that the user $u_i$ rates the item $v_j$ as $k$.
Let $U(u_i)$ denote the set of users related to user $u_i$. In addition,  $V^k(u_i)$ is defined as the set of items that the user $u_i$ rates $k$ to, while $U^k(v_j)$ as the set of users who give a rating $k$ to $v_j$.

\textbf{Problem Formulation}. Let  $\mathbf{p}_i$, $\mathbf{q}_j \in \mathbb{R}^{D}$ denote initial embeddings of the target user $u_i$ and item $v_j$. Given user-user relations and user-item interactions, the task is to predict the explicit score $\hat{r}_{ij}$ that user $u_i$ will rate item $v_j$.

\section{METHODOLOGIES}
\subsection{Overview}
Figure \ref{fig:overview} provides the overall architecture of our model. We aim to construct a heterogeneous global graph, and extract cross-semantic relations and rich topological information from it. We first build item-item subgraph $G_{vv}=\{V,\mathcal{E}_v\}$ by similarity between items. The edge $(v_j,v_m,r_v)$ in $\mathcal{E}_v$ means items $v_{j}$ and $v_{m}$ are similar. We also define $V(v_j)$ to denote the set of items similar to item $v_j$. We combine these three graphs $\{G_{uu},G_{uv},G_{vv}\}$ into 
a heterogeneous global graph $G$, which contains two kinds of nodes, three kinds of meta-paths and $K+2$ kinds of edges, i.e. user-user relation edge, $K$ kinds of rating edges, and item-item similarity edge.
In order to get a better graph structure with respect to the downstream task, capture implicit connections and filter out possible noise, we design the Graph Learner to optimize user-user (u2u) and item-item (i2i) connections. Moreover, the refined global graph is passed as input to a heterogeneous graph neural network to distill high-order complex semantic information. We employ a rating predictor to predict the score that target user will rate the candidate item. We design a hybrid loss to train our model.

\begin{figure}
    \centering
    \includegraphics[width=1.0\linewidth]{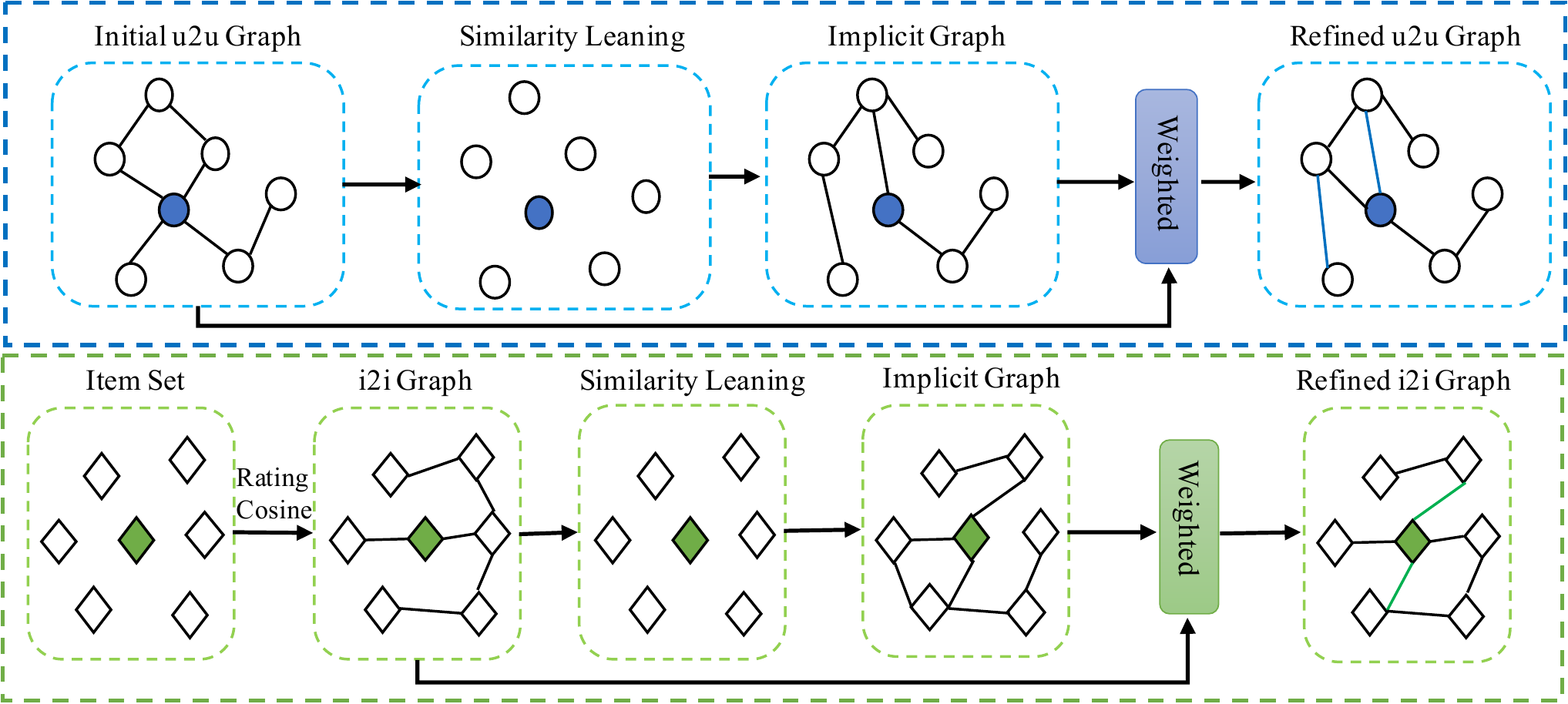}

    \caption{The structure of proposed Global Graph Learning.
    }
    \vspace{-0.5cm}
    \label{fig:GCL}
\end{figure}
\subsection{Global Graph Learning}
Figure \ref{fig:GCL} shows the architecture of Global Graph Learning module, which constructs the heterogeneous global graph and optimizes the graph structures. 
It should be noted that there is usually no connection information between items in the raw data, but item-item connections can enrich the graph structure and improve the receptive field, which allows us to extract more information of both users and items.
For this reason, we need to construct the item-item edges first. Then, we employ Graph Learner (GL) to optimize the user-user and item-item subgraph by adding or removing the edges with the method of calculating the similarity of node embeddings \cite{chen2020iterative}. We will introduce the details below.

\textbf{Item-item Connections Construction}. We utilize the rating matrix $\mathbf{R}\in \mathbb{R}^{N\times M}$ to calculate the cosine similarity between items following previous work \cite{fan2020graph}. For the rating matrix $\mathbf{R}$, we take the $j$-th column vector  $\mathbf{e}_j$ as the item $v_j$ vector. The similarity calculation formula is denoted as follows: 
\begin{equation}
\begin{aligned}
    score(v_j,v_m)=cosine(\mathbf{e}_j,\mathbf{e}_m)=\frac{\mathbf{e}_j \cdot \mathbf{e}_m}{\left \|\mathbf{e}_j \right \| \left \|\mathbf{e}_m \right \|}
\end{aligned}\label{eqn:i2i}
\end{equation}
For each item, we choose the most similar $K_I$ items to create the edges. In this way, we construct the item-item connections.


\textbf{Graph Learner}. Due to the noise or lack of possible information in the original graph structure, we propose to adapt Graph Learner to optimize the input u2u and i2i subgraphs' topologies. 

For u2u Graph Learner, the input is the initial subgraph $G_{uu}$ with node set $\{u_1,u_2,\dots,u_N\}$ and embedding set $\{\mathbf{p}_1,\mathbf{p}_2,\dots,\mathbf{p}_N\}$. For the target node $u_i$, we apply the multi perspective learning method to calculate the similarity between $u_i$ and all nodes as follows:
\begin{equation}
\begin{aligned}
    sim(u_i,u_n)=\frac{1}{F}\sum_{f=1}^F{sim^f(u_i,u_n)}, n=1,2,\dots,N
\end{aligned}\label{eqn:multiPer}
\end{equation}
Where $F$ is the number of perspectives. For each perspective, we can choose one from three methods, which are called \emph{weighted cosine},  \emph{attention}, and \emph{add attention}:
\begin{equation}
\begin{aligned}
    sim^f(u_i,u_n)=cosine(\mathbf{W}^c_f\mathbf{p}_i,\mathbf{W}^c_f\mathbf{p}_n)
\end{aligned}\label{eqn:weighted-cosine}
\end{equation}
\begin{equation}
\begin{aligned}
    sim^f(u_i,u_n)=(\mathbf{W}^a_f\mathbf{p}_i)^T(\mathbf{W}^a_f\mathbf{p}_n)
\end{aligned}\label{eqn:attention}
\end{equation}
\begin{equation}
\begin{aligned}
    sim^f(u_i,u_n)=\sigma({\mathbf{w}^d_f}^T\mathbf{p}_i+{\mathbf{w}^d_f}^T\mathbf{p}_n)
\end{aligned}\label{eqn:add_attention}
\end{equation}
Equation \ref{eqn:weighted-cosine} is the principle of \emph{weighted cosine}, and $\mathbf{W}^c_f$ is the weight of neural network. Equation \ref{eqn:attention} shows the calculation method of \emph{attention}, where $\mathbf{W}^a_f$ is a weight matrix. Equation \ref{eqn:add_attention} presents the principle of the \emph{add attention}, where $\mathbf{w}^d_f$ maps the embedding of the node to $1$ dimension and $\sigma$ is the ReLU function. During the experiment, we mainly use the \emph{weighted cosine} method, and the other two methods will be compared in the ablation study.

For all the target user nodes in one batch, the initial adjacency matrix with all nodes is $\mathbf{A}_u \in \mathbb{R}^{B\times N}$, where $B$ is the number of target user nodes in the current batch. And we can obtain a new learned implicit adjacency matrix $\mathbf{A}_u'$ with similarity calculation. Though the initial graph may be noisy or missing information, it still contains rich valuable topological information. Therefore, we employ a weight value $\lambda_w$ to combine the implicit matrix with the initial matrix:
\begin{equation}
\begin{aligned}
    \tilde{\mathbf{A}}_u=\lambda_w\mathbf{A}'_u+(1-\lambda_w)\mathbf{A}_u
\end{aligned}\label{eqn:ad_attention}
\end{equation}
Each element in the refined matrix $\tilde{\mathbf{A}}_u$ represents the similarity of two nodes. In order to prevent information redundancy caused by too many edges, we set a truncation length $L$. For each target node, we truncate the first $L$ nodes with the highest similarity to establish new connections, and the remaining nodes are not connected to the target node. In this way, we can get the refined subgraph $G'_{uu}$.



For the input i2i subgraph $G_{vv}$, we can apply the same method to get the refined subgraph $G'_{vv}$.

\textbf{Anchor-based Graph Learner}. In the real world, the number of nodes is often very huge. For target nodes, if we calculate the similarity of all the nodes to them, the costs of computation are high.
Inspired by \cite{chen2020iterative}, we proposed a scalable Anchor-based Graph Learner (AGL). Next, we take the item-item subgraph as an example. For target item nodes, we randomly select $H_i (H_v\ll M)$ nodes as the anchor nodes set $\{v_{m_1},v_{m_2},\dots,v_{m_{H_v}}\}$.  We can get the initial adjacency matrix between target nodes and anchor nodes $\mathbf{A}_{v,anchor} \in \mathbb{R}^{B\times H_v}$ from the initial connections. We calculate the target-anchor similarity matrix $\mathbf{A}'_{v,anchor}$, as Equation \ref{eqn:multiPer}. Then we use the weight value $\lambda_w$ and the truncation length $L$ to calculate the refined item-item subgraph $G'_{vv}$. Similarly, we can randomly select $H_u(H_u\ll N)$ user nodes as the anchor nodes set, and employ AGL to get the refined user-user subgraph $G'_{uu}$.


\subsection{Heterogeneous Graph Neural Network}
In this subsection, we discuss how to extract user and item latent features in a unified perspective, based on the refined global graph $G'=\{G'_{uu},G_{uv},G'_{vv}\}$ learned by Global Graph Learning. The global graph includes three kinds of semantic meta-paths: user-user relations, user-item interactions and item-item similarities. Inspired by \cite{schlichtkrull2018modeling}, we employ a Heterogeneous Graph Neural Network (HGNN) to extract high-order information and fuse different semantic information.

We employ $T$-layer HGNN to model our refined heterogeneous global graph to distill cross-semantic information. For the target user $u_i$ and target item $v_j$, the initial input embeddings of the first layer are $\mathbf{p}^{(0)}_i=\mathbf{p}_i$ and $\mathbf{q}^{(0)}_j=\mathbf{q}_j$. Let $\mathbf{p}^{(t)}_i$ and $\mathbf{q}^{(t)}_j$ denote the representations of user $u_i$ and the item $v_j$ after the propagation of $t$-th layer. We next introduce the user node aggregation and item node aggregation in each layer of HGNN.

\textbf{User node aggregation}. 
Generally, for each user node in the refined heterogeneous global graph, there exits one type of edges $r_u$ connecting the user neighbors and $K$ types of edges $r_k, (k\in\{1,2,\dots,K\})$ connecting the item neighbors. For the user-user social semantic connections, we aggregate the features of user neighbors as follows:

\begin{equation}
\begin{aligned}
    \mathbf{p}^{(t+1)}_{i,u}=\mathbf{b}_u^{(t+1)}+\sum_{u_n\in U(u_i)}{\frac{1}{c_{i,n}}}\mathbf{W}^{(t+1)}_u\mathbf{p}^{(t)}_n
\end{aligned}\label{eqn:grad}
\end{equation}
\begin{equation}
\begin{aligned}
    c_{i,n}= \sqrt{\left|U(u_i) \right| \left|U(u_n)\right|} 
\end{aligned}\label{eqn:grad}
\end{equation}
Where $\mathbf{W}^{(t+1)}_u$ is a trainable transformation matrix, $\mathbf{b}_u^{(t+1)}$ is the bias vector, and $c_{i,n}$ is the normalization coefficient.

Similarly, we perform user $u_i$ node aggregation based on $K$ types of user-item rating connections. Specifically, for each type of edges $r_k$, we also aggregate neighbor items under the same rating level as follows
\begin{equation}
\begin{aligned}
    \mathbf{p}^{(t+1)}_{i,k}=\mathbf{b}_k^{(t+1)}+\sum_{v_m\in V^k(u_i)}{\frac{1}{c_{i,m}}}\mathbf{W}^{(t+1)}_k\mathbf{q}^{(t)}_m
\end{aligned}\label{eqn:grad}
\end{equation}

\begin{equation}
\begin{aligned}
    c_{i,m}= \sqrt{\left|V^k(u_i) \right| \left|U^k(v_m)\right|} 
\end{aligned}\label{eqn:grad}
\end{equation}
where $k \in \{1,2,\dots,K\}$.

For user $u_i$, we accumulate all messages propagated by different $K+1$ types of edges $[ \mathbf{p}^{(t+1)}_{i,u}, \mathbf{p}^{(t+1)}_{i,1},\dots, \mathbf{p}^{(t+1)}_{i,K}]$. Then, we aggregate the information of these $K+1$ embeddings:
\begin{equation}
\begin{aligned}
    \mathbf{p}^{(t+1)}_i=\sigma(\frac{1}{K+1}(\mathbf{p}^{(t+1)}_{i,u}+\sum_{k=1}^K{\mathbf{p}^{(t+1)}_{i,k}}))
\end{aligned}\label{eqn:grad}
\end{equation}
$\sigma$ is the ReLU function, $\mathbf{p}^{(t+1)}_i$ is the output embedding of user $u_i$ in $t+1$-th HGNN layer. It is worth noting that, for the current layer, we integrate two kinds of meta-paths (user-user, user-item) information into the user's features, while the item features already contain the item-item semantic information after $t$ layers aggregation. Therefore, the user's features can also fuse item-item semantic information by the multi-layer HGNN.

\textbf{Item node aggregation}.
The target item $v_j$ also involves in two meta-paths: item-item similarity and user-item interactions including $K$ types of edges. 
Similarly, for the $t+1$-th layer, we propagate different mesages from $K+1$ types of edges and obtain $K+1$ embeddings $[\mathbf{q}^{(t+1)}_{j,v}, \mathbf{q}^{(t+1)}_{j,1},\dots, \mathbf{q}^{(t+1)}_{j,K}]$ of $v_j$. Then we aggregate these embeddings into the 
output embedding $\mathbf{q}^{(t+1)}_j$:

\begin{equation}
\begin{aligned}
    \mathbf{q}^{(t+1)}_j=\sigma(\frac{1}{K+1}(\mathbf{q}^{(t+1)}_{j,v}+\sum_{k=1}^K{\mathbf{q}^{(t+1)}_{j,k}}))
\end{aligned}\label{eqn:grad}
\end{equation}

After $T$ layers of HGNN, we can extract high-order and cross-semantic information from the refined heterogeneous global graph, which enables us to distill more latent features of users and items.
\subsection{Rating Predictor}
The initial embeddings and output of each HGNN layer constitute the user $u_i$ embedding lists $[\mathbf{p}^{(0)}_i,\mathbf{p}^{(1)}_i,\dots,\mathbf{p}^{(T)}_i] $ and item $v_j$ embedding lists $[\mathbf{q}^{(0)}_j,\mathbf{q}^{(1)}_j,\dots,\mathbf{q}^{(T)}_j]$. In this module, we design the shared attention mechanism to get the final user and item latent embeddings. For the user $u_i$, the final embedding is defined as follows:
\begin{equation}
\begin{aligned}
    \mathbf{p}_i^*=\sum_{t=0}^T\alpha_{t,u}\mathbf{p}^{(t)}_i
\end{aligned}\label{eqn:attention_u}
\end{equation}
\begin{equation}
\begin{aligned}
    \alpha_{t,u}^*=\mathbf{s}^T\sigma(\mathbf{W}_s\mathbf{p}_i^{(t)}+\mathbf{b}_s)
\end{aligned}\label{eqn:softmax alpha_u}
\end{equation}
\begin{equation}
\begin{aligned}
    \alpha_{t,u}=\frac{\alpha_{t,u}^*}{\sum_{t'=0}^T\alpha_{t',u}^*}
\end{aligned}\label{eqn:alpha_u}
\end{equation}
Where $\mathbf{W}_s$, $\mathbf{s}$ and $\mathbf{b}_s$ are the shared trainable parameters, $\sigma$ is the ReLU function. And the final embedding $\mathbf{q}_j^*$ of item $v_j$ can be calculated in the same way. In this paper, we focus on the rating prediction task in social recommendation, so we design the predictor based on multi layer perceptron (MLP) :
\begin{equation}
\begin{aligned}
    \hat{r}_{ij}=\text{MLP}([\mathbf{p}_i^*,\mathbf{q}_j^*])
\end{aligned}\label{eqn:prediction}
\end{equation}
Where $[,]$ is the concatenation operation.

\subsection{Model Training}
To better train our model, we design a special loss function, which contains two aspects of loss: (i) Graph Learner loss, (ii) rating loss.

\textbf{Graph Learner Loss}. In our work, the updated graph structure plays an important role in rating prediction. In order to obtain the better graph topology with respect to the social recommendation task, we design the Graph Learner (GL) loss through graph regularization \cite{belkin2001laplacian,chen2020iterative}. For the u2u GL, we can get the refined adjacency matrix $\tilde{\mathbf{A}}_u\in \mathbb{R}^{B\times N}$. Generally, graph regularization is often applicable for symmetric adjacency matrix. Since $\tilde{\mathbf{A}}_u$ is not symmetric, we first transform it to be symmetric as follows:
\begin{equation}
\begin{aligned}
    \hat{\mathbf{A}}_u=\tilde{\mathbf{A}}_u\mathbf{\Delta}^{-1}\tilde{\mathbf{A}}_u^T
\end{aligned}\label{eqn:graph regularization}
\end{equation}
Where $\mathbf{\Delta}\in\mathbb{R}^{N\times N}$ ($\Delta_{ii}=\sum_{k=1}^NA_{u,ki}$) is a diagonal matrix. As we all know, that values change smoothly among adjacent nodes is a widely applied assumption. Therefore, we utilize $\hat{\mathbf{A}}_u$ and initial user feature matrix $\mathbf{P}$ to design the smoothness loss as follows:
\begin{equation}
\begin{aligned}
    L(\hat{\mathbf{A}}_u,\mathbf{P})=\frac{1}{2B^2}\sum_{i,n}\hat{A}_{u,in}\left \| \mathbf{p}_i-\mathbf{p}_n\right \|^2=\frac{1}{B^2}\mathbf{tr}(\mathbf{P}^T\mathbf{L}\mathbf{P})
\end{aligned}\label{eqn:smoothness}
\end{equation}
Where $\mathbf{tr}(\cdot)$ indicates the trace of a matrix, $\mathbf{L}=\mathbf{D}_u-\hat{\mathbf{A}}_u$ is the graph Laplacian, and $\mathbf{D}_u=\sum_n\hat{A}_{u,in}$ denotes the
degree matrix. However, only minimizing the smoothness loss will cause over smoothing, so we impose constraints\cite{chen2020iterative} to control smoothness as follows:
\begin{equation}
\begin{aligned}
    C(\hat{\mathbf{A}}_u)=-\frac{\beta_1}{B}\mathbf{1}^Tlog(\hat{\mathbf{A}}_u\mathbf{1})+\frac{\beta_2}{B^2}\left \| \hat{\mathbf{A}}_u \right\|_2
\end{aligned}\label{eqn:constraints}
\end{equation}
Where $\mathbf{1}$ indicates the vector in which elements are $1$, and $\left \| \hat{\mathbf{A}}_u \right\|_2$ indicates the Euclidean norm of $\hat{\mathbf{A}}_u$. We then define the overall Graph Learner loss of u2u GL as the sum of the previously defined losses:

\begin{equation}
\begin{aligned}
    L^u_{G}=\beta L(\hat{\mathbf{A}}_u,\mathbf{P})+ C(\hat{\mathbf{A}}_u)
\end{aligned}\label{eqn:graph learner loss}
\end{equation}
$\beta$ is a non-negative hyper-parameters. 

While for u2u AGL, we can convert the refined adjacency  matrix $\tilde{\mathbf{A}}_{u,anchor}\in \mathbb{R}^{B\times H_u}$ to the symmetric matrix $\hat{\mathbf{A}}_{u,anchor}$ as Equation \ref{eqn:graph regularization}. And we can rewrite Equation \ref{eqn:graph learner loss} to define the Anchor-based Graph Learner loss:
\begin{equation}
\begin{aligned}
    L^u_{G}=\beta L(\hat{\mathbf{A}}_{u,anchor},\mathbf{P})+ C(\hat{\mathbf{A}}_{u,anchor})
\end{aligned}\label{eqn:anchor graph learner loss}
\end{equation}

We can also calculate i2i GL or AGL loss $L^v_{G}$ by the same method.

\textbf{Rating Loss}.
For the task of rating prediction, we adopt mean square error (MSE) loss function as:
\begin{equation}
\begin{aligned}
    L_{r}=\frac{1}{B}\sum_{i,j}\left \|\hat{r}_{ij}-r_{ij} \right \|^2
\end{aligned}\label{eqn:ratingloss}
\end{equation}
Where $r_{ij}$ is the ground-truth value. For our model, we apply a hybrid loss to jointly learn the parameters:
\begin{equation}
\begin{aligned}
    L=L_{r}+\gamma_u L_{G}^u+\gamma_v L_{G}^v+\eta \Omega(\Theta)
\end{aligned}\label{eqn:loss}
\end{equation}
$\gamma_u$, $\gamma_v$ and $\lambda$ are non-negative hyper-parameters. $\Theta$ is the trainable parameters, $\Omega(\cdot)$ denotes the L2 regularization. Through multiple epochs of optimization, we can iteratively learn an optimized global graph structure with respect to the social recommendation as well as reliable user and item features.
\subsection{Model Complexity Analysis}
\textbf{GL-HGNN.} As for GL-HGNN, the computational cost of the Graph Learner is $\mathcal{O}(E(N+M)D)$ for $N$ user nodes, $M$ item nodes and $E$ missing user-item rates to be predicted. The computational cost of HGNN is $\mathcal{O}(TX(M+N)D)$, where $T$ denotes the number of layers and $X$ indicates the average neighbors of each node. The rating task costs $\mathcal{O}(EdD)$ where $d$ is the hidden size, while the computational complexity of the hybrid loss is $\mathcal{O}(E(N+M)D)$. The overall cost is about $\mathcal{O}((TX+E)(N+M)D+EdD)$. If we assume that $E\approx N+M$ and $TX, d\ll N+M$, the overall time complexity is $\mathcal{O}((N+M)^2D)$. 

\textbf{AGL-HGNN.} As for AGL-HGNN, the computational cost of the Anchor-based Graph Learner is $\mathcal{O}(E(H_u+H_v)D)$, while computing node embeddings by HGNN costs $\mathcal{O}(TX'(M+N)D)$, where $X'$ indicates the average neighbors of each node. The rating task also costs $\mathcal{O}(EdD)$, and computing the hybrid loss costs $\mathcal{O}(E(H_u+H_v)D)$. As $H_u,H_v,d\ll(N+M)$, the overall time complexity of AGL-HGNN is $\mathcal{O}(TX'(N+M)D)$, which is linear to the number of user and item. Therefore, 
AGL-HGNN can significantly reduce the computational complexity.

\begin{table}[htbp]
    \caption{Statistics of datasets. }
    \label{tab:ws-darts}
    \centering
    \small
    \begin{tabular}{lcccc}
    \toprule
    \textbf{Dataset}   & \textbf{Ciao-5} & \textbf{Ciao-28} & \textbf{Epinions}& \textbf{Flixster}\\
    \midrule
    $\#$ of Users & 2,248 &10,994&22,164&147,612\\
    $\#$ of Items  & 16,861 &112,802&296,277& 48,794\\
    $\#$ of Ratings & 36,065 & 304,493&922,267&8,196,067\\
    $\#$ of Ratings Density & 0.095$\%$& 0.025$\%$&0.014$\%$&0.114$\%$\\
    Rating Range & [1,5]& [1,5]&[1,5]&[0.5,5]\\
    \midrule
    $\#$ of Links & 52,907 &131,427&362,433&2,442,886\\
    $\#$ of Links Density& 1.047$\%$ & 0.108$\%$ &0.073$\%$&0.011$\%$\\
    \bottomrule
    \end{tabular}
\end{table}

%% file: 2-exp.tex
\section{Experiments}

In this section, we will detail the settings of our experiment and present the experimental results\footnote{Our code and data will be released for research purpose.}. To fully demonstrate the superiority of our model, we conduct experiments to verify the following four research questions (RQ):

\begin{itemize}
    \item \textbf{(RQ1)}: Compared with the state-of-the-art models, does our model achieve better performance?
    
    \item \textbf{(RQ2)}: What are the impacts of key components on model performance?
    
    \item \textbf{(RQ3)}: How does the setting of hyper-parameters (such as the truncation length in Graph Learner) affect our model?
    
    \item \textbf{(RQ4)}: How can Global Graph Leaning module improve the performance of our model?
\end{itemize}

\begin{table*}[t]
    \centering
    \caption{Performance comparison of different models on the four datasets. The smaller the RMSE and MAE, the better the performance.}
    \label{results}
    \begin{tabular}{p{2.5cm}<{\centering}p{1.0cm}<{\centering}p{1.0cm}<{\centering}p{1.0cm}<{\centering}p{1.0cm}<{\centering}p{1.0cm}<{\centering}p{1.0cm}<{\centering}<{\centering}p{1.0cm}<{\centering}p{1.0cm}<{\centering}}
    \toprule
    \multirow{2}{*}{\bfseries Models }&\multicolumn{2}{c}{\bfseries Ciao-5 }&\multicolumn{2}{c}{\bfseries Ciao-28 }&\multicolumn{2}{c}{\bfseries Epinions }
    &\multicolumn{2}{c}{\bfseries Flixster }\\
    &RMSE&MAE&RMSE&MAE&RMSE&MAE&RMSE&MAE\\
    \midrule
    SocRec&1.0288&0.7847&1.1881&0.8571&1.1964&0.9045&1.1237&0.8323\\
    TrustMF&1.0182&0.8004&1.1506&0.8799&1.1723&0.8832&1.0594&0.8132\\
    TrustSVD&0.9796&0.7845&1.0986&0.8480&1.1394&0.8601&1.0402&0.8097\\
    DSCF&0.9785&0.7651&1.0932&0.8391&1.1295&0.8532&1.0220&0.8034\\
    \midrule
    GC-MC&0.9260&0.7230&1.0736&08226&1.1168&0.8594&0.9870& 0.7571\\
    GraphRec&0.9226&0.7006&1.0503&0.8157&1.1036&0.8485&0.9360&0.7161\\
    DANSER&{0.9038}&{ 0.6857}&1.0496&0.8102&{ 1.0821}&{0.8164}&0.9400&0.7113\\
    GraphRec+&0.9191&0.7065&{ 1.0477}&{ 0.8088}&1.0943&0.8377&0.9225&0.7054\\
    \midrule
    {\bfseries GL-HGNN}&{\bfseries 0.8615} & \bfseries 0.6497& {\bfseries 1.0320} &{ 0.7763}&{\bfseries 1.0709}&{ \bfseries 0.8017}&0.9142&0.6993\\
    {\bfseries AGL-HGNN }&0.8676&0.6535& {1.0330 } &{\bfseries 0.7758}&{1.0727}&{0.8092 }&{\bfseries 0.9091}&{\bfseries 0.6899}\\
    \bottomrule
    \end{tabular}
\end{table*}
\subsection{Experiment Setup}

\subsubsection{Datasets}\label{sec:standalone}
We conduct experiments on several public social recommendation benchmark datasets \emph{Ciao}\footnote{http://www.ciao.co.uk} \cite{tang2012etrust}, \emph{Epinions}\footnote{http://www.epinions.com} \cite{tang2012mtrust} and \emph{Flixster}\footnote{https://www.flixster.com}\cite{jamali2010matrix}, which all contain rating information and social networks. The detailed statistics of dataset are given in \autoref{tab:ws-darts}.
\begin{itemize}
    \item \textbf{Ciao}: \emph{Ciao} is drieved from a popular social networking e-commerce platform. We process two available versions of the Ciao datasets, separately called Ciao-5 and Ciao-28. Ciao-5 collects 5 categories of items and their corresponding users, while Ciao-28 contains all 28 categories of items (such as DVDS) and users. The rating range is $[1,5]$ with the step size $1$..
    \item \textbf{Epinions}: \emph{Epinions} comes from a social based product review platform. The rating values contain five discrete numbers, which are $\{1,2,3,4,5\}$.
    \item \textbf{Flixster}: 
    \emph{Flixster} comes from a popular movie review website, where people can add others as friends to create the social network. The range of rating value is $[0.5,5]$ with the step size $0.5$.
\end{itemize}

For each dataset, we select $20\%$ as the test set, $10\%$ as valid set and remaining $70\%$ as training set.

\subsubsection{Evaluation Metrics}
In order to better evaluate the performance of models, we employ two widely used metrics, namely RMSE (root mean square error) and MAE (mean absolute error) \cite{wang2018exploring}. The two metrics both indicate the error between the predicted value and the ground-truth, while RMSE is more sensitive to outliers.

    

\subsubsection{Baselines}
To evaluate the performance of our model, we select representative seven models, including classic and state-of-the-art (SOTA) social recommendation models as follows:
\begin{itemize}
    \item \textbf{SoRec} \cite{ma2008sorec}: It learns users' feature vectors by decomposing the scoring matrix and the social relation matrix simultaneously.
    \item \textbf{TrustMF} \cite{10.5555/2540128.2540524}: According to the direction of trust, this model maps users to the trusted space and the trustee space, by matrix factorization.
    \item \textbf{TrustSVD} \cite{guo2015trustsvd}: This is one matrix factorization-based model, aggregating friends embeddings into target users embeddings to learn explicit and implicit information.
    \item \textbf{DSCF} \cite{fan2019deep}: This method proposes a deep learning-based framework, which captures the influence of distant social relationships on target users. 
    \item \textbf{GC-MC}\cite{berg2017graph}: This model generates the implicit information between users and items in the form of information transfer in the bipartite interaction graph. However, it only models the links between users and item. In the experiment, we also join social network to make predictions.
    \item \textbf{GraphRec} \cite{fan2019graph}: This method jointly captures the user-item interaction and opinion between users and items from user-item graph, and learns the heterogeneous social relationship between users from user-user graph.
    \item \textbf{DANSER} \cite{wu2019dual}: This method constructs a large graph that contains user-user, item-item, and user-item sub-graphs. By modeling this large graph, it learns the dynamic and static attributes of users and items, and then fuses the dual attributes to predict users' ratings on target items through one fusion strategy.
    \item \textbf{GraphRec+} \cite{fan2020graph}: On the basis of Graphrec, Graphrec+ not only models user-item and user-user graphs, but adds item-item graph to aggregate information between similar items.
    
\end{itemize}

\subsubsection{Parameters Setting}
We implement our model based on Pytorch and DGL. We set the embedding dimension $D=64$, and the batch size as $128$. For all trainable parameters, we initialize them with a Gaussian distribution with an average of $0$ and a standard deviation of $0.01$. We use mini-batch Adam optimizer to train the model parameters with initial learning rate of $0.001$. In order to prevent over-fitting, we add dropout layers with a probability value of $0.4$ during training. In construction of item-item edges, we select top $20$ items for each item to build connections, according to the similarity cosine values. For the Graph Learner, we search the weight $\lambda_w$ of learned implicit graph structure in $ [ 0.1,0.3,0.5,0.7,0.9 ] $. The number $F$ of perspectives of node similarity calculation in the Graph Learner, is tuned in the set of $[ 1,2,3,4 ]$.  For the truncation length $L$ in Graph Learner, we obtain the optimal value in the range $[ 20,40,60,80,100 ]$ through the grid search. In addition, we set the number of graph neural network layers in range of $[1,2,3,4]$. 

In addition, we also apply Anchor-based Graph Learner module in our experiments. We define the anchor rate $\tau=H_u/N=H_v/M$. We test the value of $\tau$ in the set  $[0.01,0.02,0.05,$ $0.1,0.15,0.2]$.             

For all the baselines, in order to achieve the best performance of these models, we set the parameters strictly according to the papers.
\begin{figure*}[t]
    \centering
    \begin{subfigure}{0.32\linewidth}
        \includegraphics[width=\textwidth]{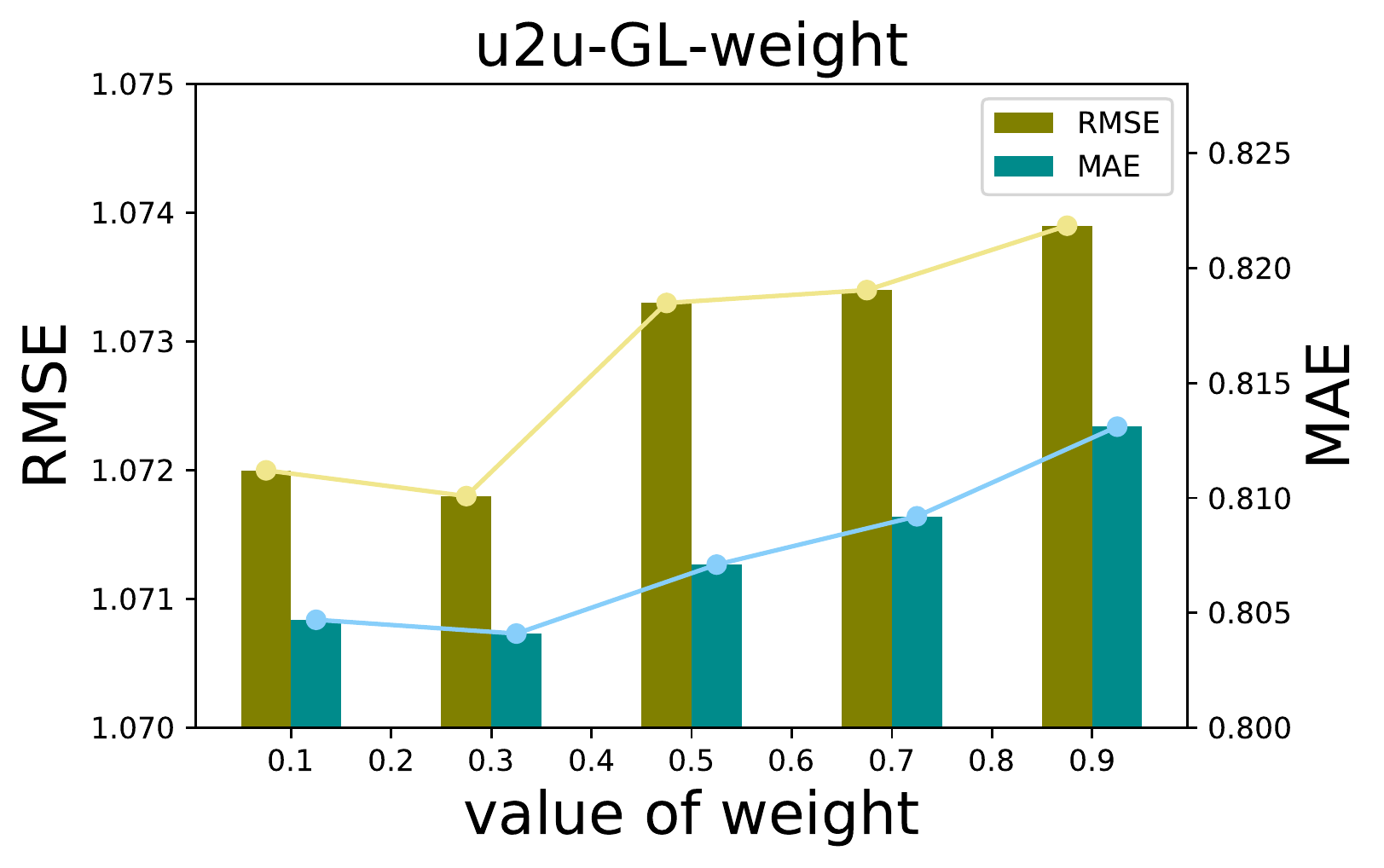}
        \includegraphics[width=\textwidth]{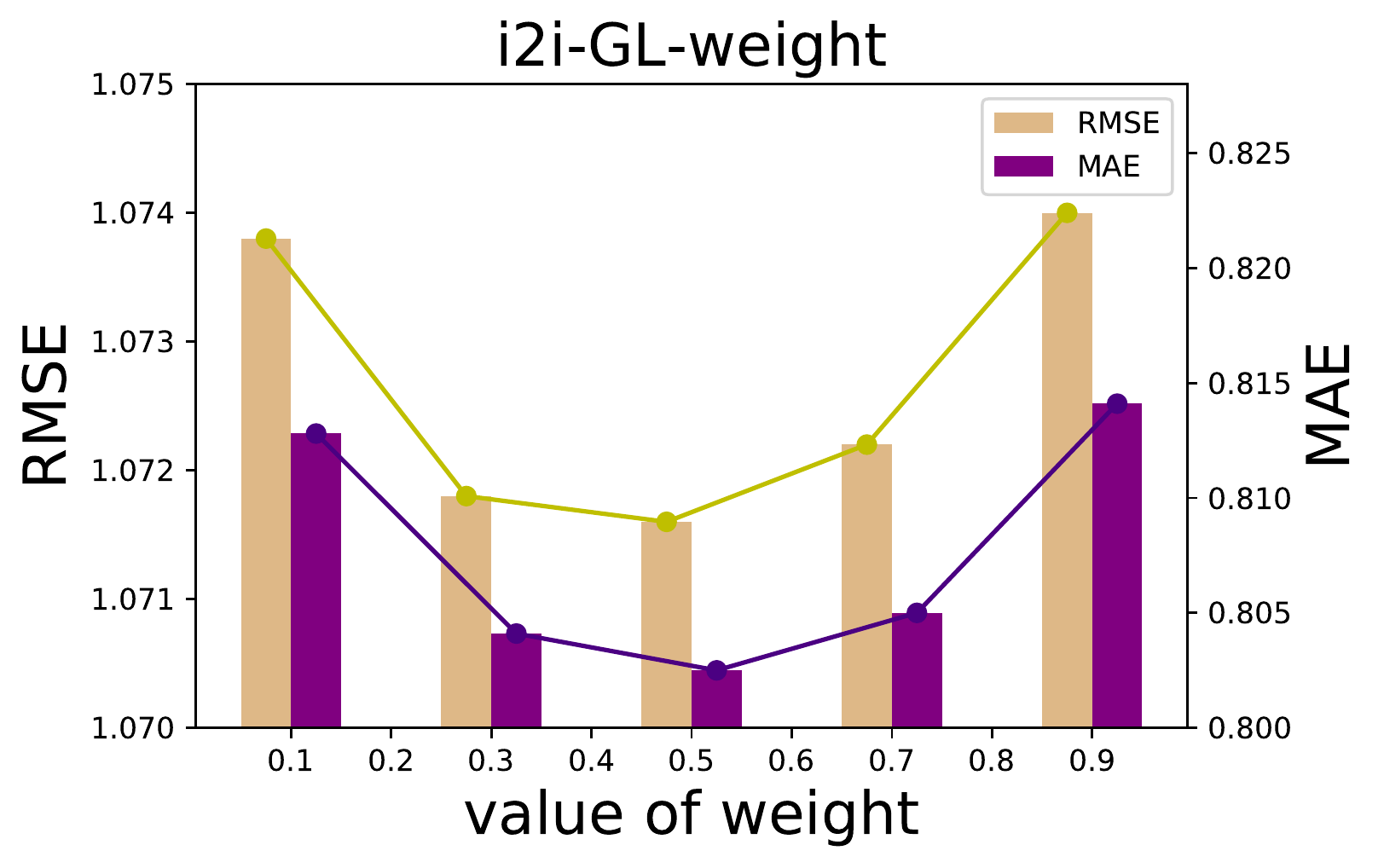}
        \caption{Graph Learner-Weight}\label{fig:search101}
    \end{subfigure}
    \begin{subfigure}{0.32\linewidth}
        \includegraphics[width=\textwidth]{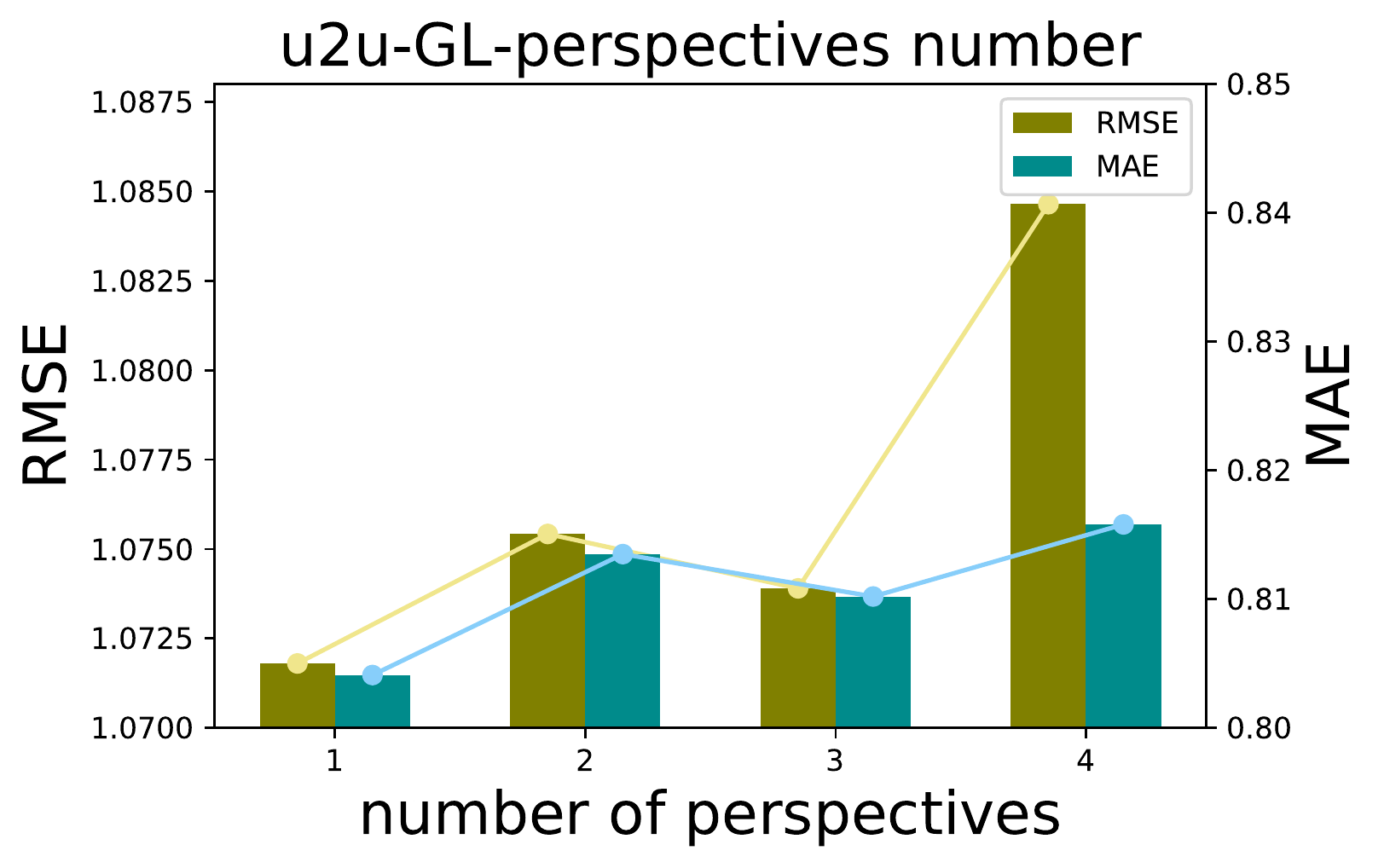}
        \includegraphics[width=\textwidth]{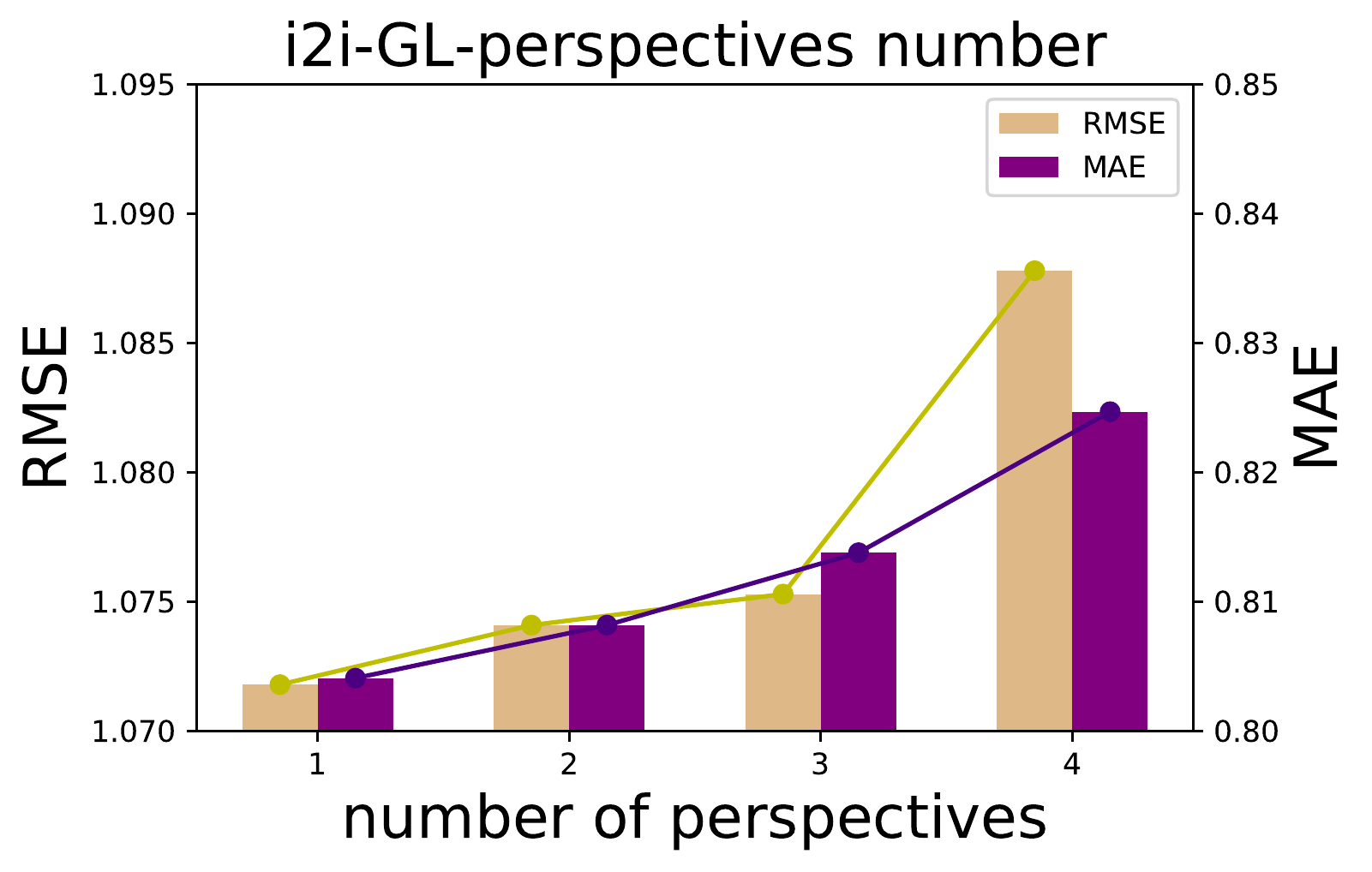}
        \caption{Graph Learner-Perspective}\label{fig:search102}
    \end{subfigure}
    \begin{subfigure}{0.32\linewidth}
        \includegraphics[width=\textwidth]{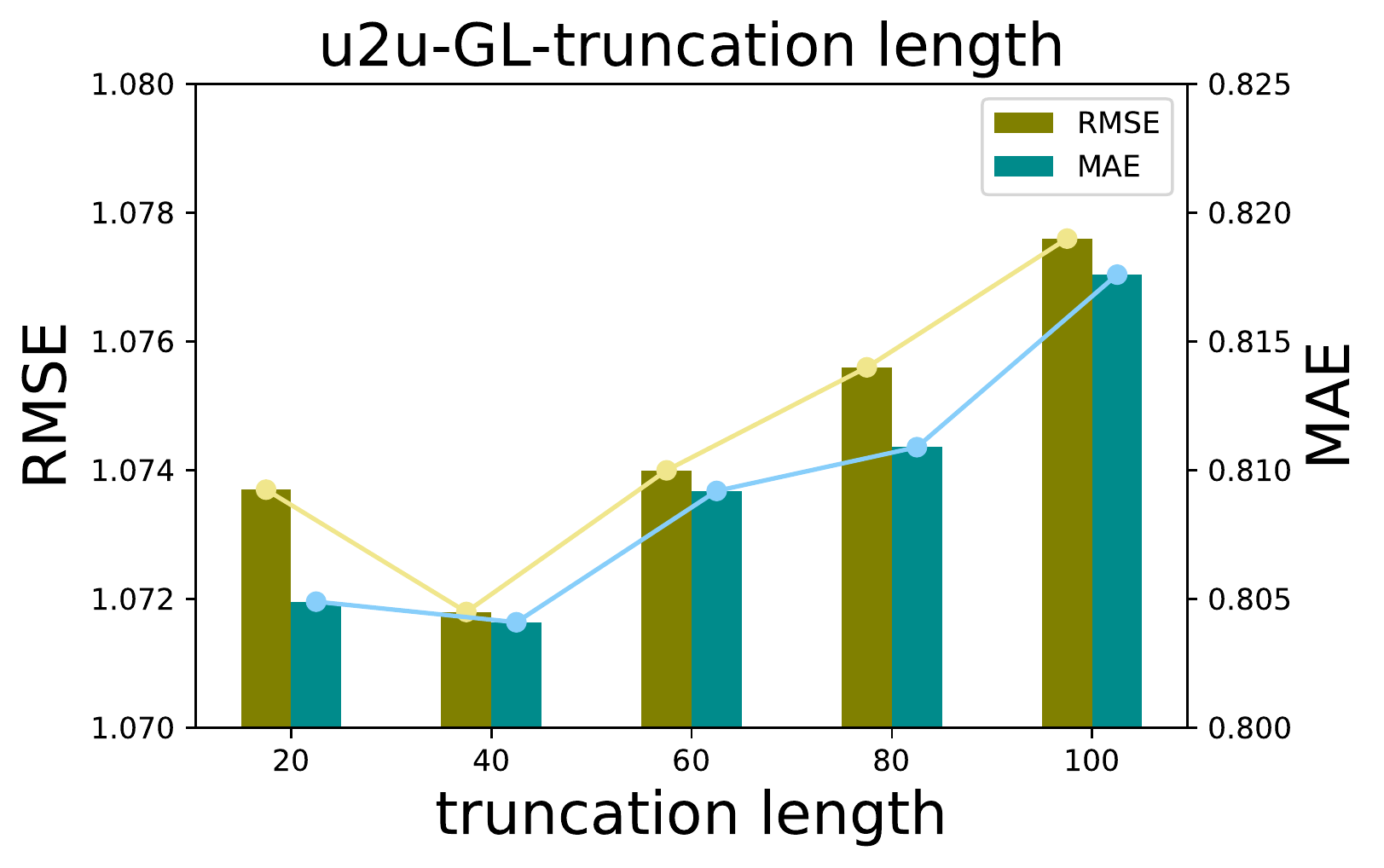}
        \includegraphics[width=\textwidth]{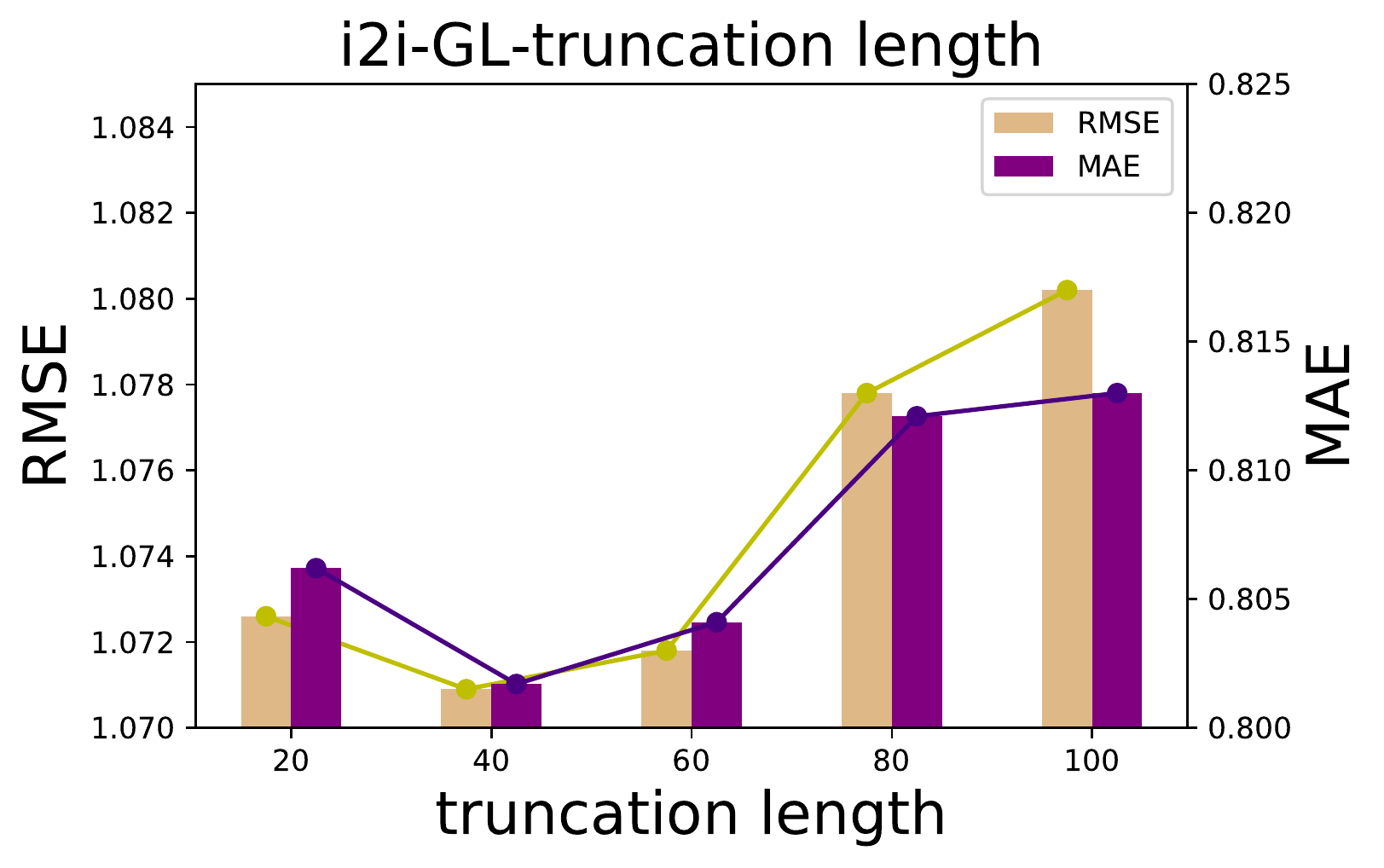}
        \caption{Graph Learner-Truncation Length}\label{fig:search103}
    \end{subfigure}
    \caption{Comparisons of different hyper-parameters w.r.t. the weight value $\lambda_w$ of learned graph structure, the number of perspectives $P$ and the truncation length $L$.}
    \label{fig:parameters}
\end{figure*}

\subsection{Results:RQ1}
The experimental results of the baseline models and our models on four datasets are shown in \autoref{results}. Based on the comparison in the table, we can summarize our findings as follows:
\begin{itemize}
    \item  Our model GL-HGNN comprehensively outperforms all the baseline models on the four datasets. The results indicate that our model is effective to the rating prediction task of the social recommendation. Different from the SOTA methods: GraphRec+ and DANSER, our approach models the heterogeneous global graph to capture high-order features and different semantic information. In addition, to obtain a better graph structure for social recommendation, GL-HGNN employs Graph Learners to optimize initial u2u and i2i connections. Besides, compared with GL-HGNN, AGL-HGNN can achieve comparable results, even better ones sometimes.

    

    \item Among all the baselines, the performance of deep learning-based methods is better than that of traditional methods, which shows that deep learning-based methods have a stronger learning ability for user relations and user-item interaction signals. Moreover, the GNN-based models achieve better results than other models without graph structure. It proves the effectiveness of GNN for social recommendation. Furthermore, GraphRec+ and DANSER achieve a better performance than other model without i2i subgraph construction. That suggests that adding extra i2i connections into the user-item graph can be helpful for social recommendation.

\end{itemize}

\begin{table}[t]
    \caption{Results of the Ablation Study. }
    \label{tab:ablation_study}
    \centering
    \small
    \begin{tabular}{lcccc}
    \toprule
    \multirow{2}{*}{\bfseries Models }& \multicolumn{2}{c}{\textbf{Ciao-5}}& \multicolumn{2}{c}{\textbf{Epinions}}\\
    &RMSE&MAE&RMSE&MAE\\
    \midrule
    \textbf{GL-HGNN}&\textbf{0.8615}&\textbf{0.6497}& \textbf{1.0709}&\textbf{0.8017}\\
    \midrule
    GL-HGNN-Attention&0.8642&0.6539&1.0745&0.8059\\
    GL-HGNN-Add Attention&0.8692&0.6531&1.0738&0.8065	\\
     \midrule
    GL-HGNN-w/o u2u GL&0.8708&0.6889&1.0746&0.8164\\
    GL-HGNN-w/o i2i GL&0.8654&0.6632&1.0764&0.8212\\
    GL-HGNN-w/o GLs&0.8813&0.6893&1.0795&0.8258\\
    GL-HGNN-w/o GLs\&i2i edges & 0.9171&0.7077&1.0860&0.8369\\
   
    \bottomrule
    \end{tabular}
\end{table}

\subsection{Ablation Study: RQ2}
In order to verify the effectiveness of some key modules, we conduct a series of ablation experiments on the Ciao-5 and Epinions datasets. The results are shown in Table \ref{tab:ablation_study}.
Firstly, we compare different calculation methods of nodes similarity in the Graph learner by replacing \emph{weighted cosine} with \emph{attention} and \emph{add attention}. As can be seen in Table \ref{tab:ablation_study}, it is clear that the \emph{weighted cosine} method is the best one of three methods to capture similar attributes between nodes. 

Besides, we explore to evaluate the effectiveness of the critical modules of GL-HGNN. We delete each module of GL-HGNN to observe the change of model performance, e.g., removing the u2u GL module and removing the i2i GL module. We can observe that the Graph Learner module is pivotal for the model performance by seeing "GL-HGNN-w/o GLs". These results demonstrate that a more suitable graph structure with respect to the downstream task plays an important role. In addition, without i2i connection information, the model performance declines to a certain extent, which shows that capturing implicit item relations from  the user rating matrix is valuable for the rating prediction task.



\begin{table}[t]
    \caption{Performance with different number $T$ of HGNN layers. }
    \label{tab:HRGCN}
    \centering
    \begin{tabular}{lcccc}
    \toprule
    \multirow{2}{*}{\bfseries Models}& \multicolumn{2}{c}{\textbf{Ciao-28}}& \multicolumn{2}{c}{\textbf{Epinions}}\\
    &RMSE&MAE&RMSE&MAE\\
    \midrule
    \textbf{GL-HGNN-1}&1.0501&0.8072&1.1276&0.8617\\
    \textbf{GL-HGNN-2}&1.0343&\textbf{0.7753}&1.0912& 0.8345\\
    \textbf{GL-HGNN-3}&\textbf{1.0320}&0.7763&\textbf{1.0709}&\textbf{0.8017}\\
    \textbf{GL-HGNN-4}&1.0398&0.7847&1.0846&0.8204\\
    \bottomrule
    \end{tabular}
\end{table}
\begin{figure}[t]
    \centering
    \begin{subfigure}{0.48\linewidth}
        \includegraphics[width=\textwidth]{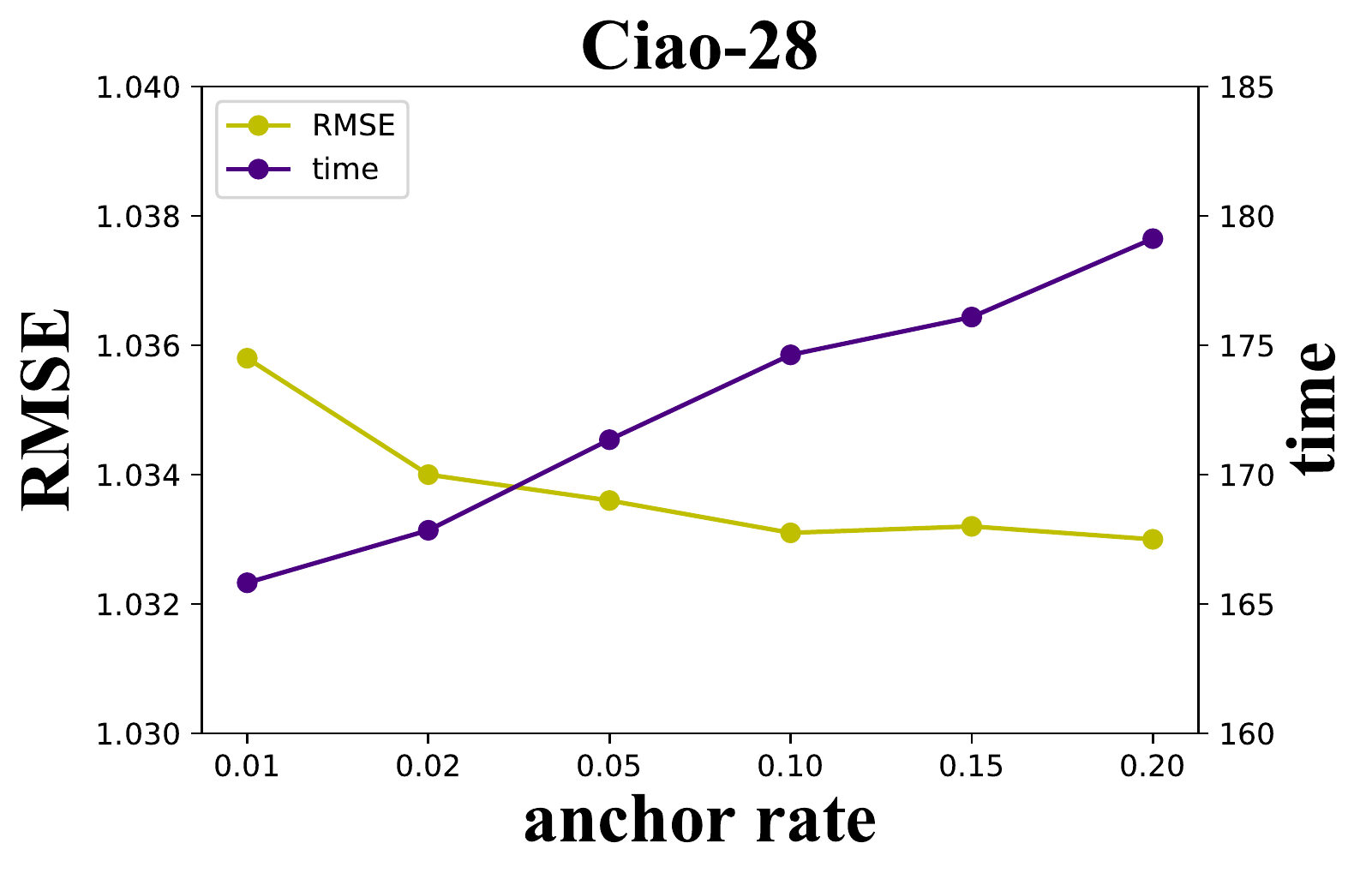}
    \end{subfigure}
    \begin{subfigure}{0.48\linewidth}
        \includegraphics[width=\textwidth]{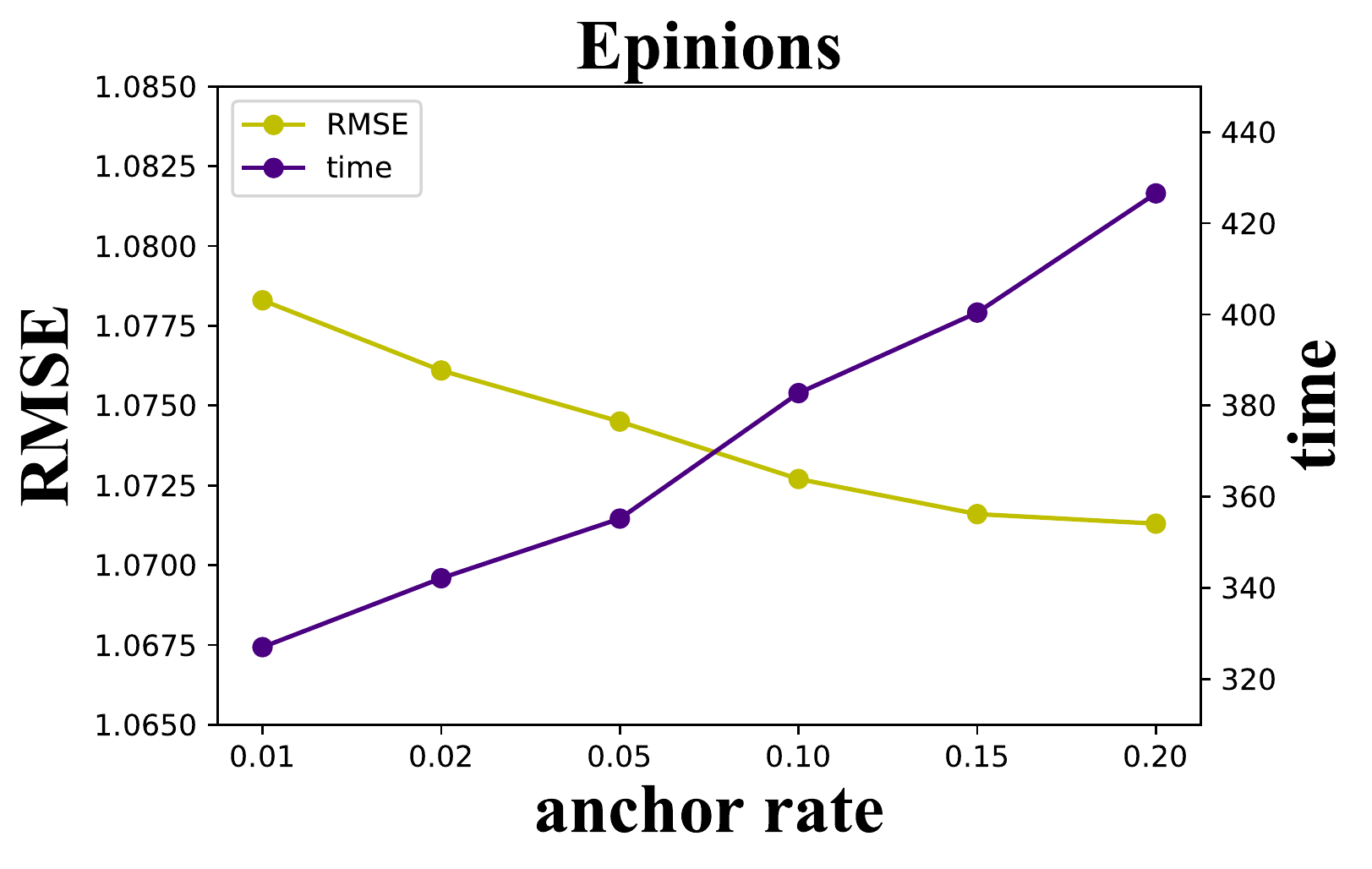}
    \end{subfigure}

    \caption{Performance comparison and running time (seconds) with different anchor rates.}
    \label{fig:anchor}
\end{figure}
\subsection{Parameter Sensitivity: RQ3}
\textbf{Global Graph Learning}
The performance of the Global Graph Learning is mainly affected by four important parameters, i.e., the weight value $\lambda_w$ of learnt implicit graph structure, the number of perspectives $P$, the truncation length $L$ and the anchor rate $\tau$. 
\begin{itemize}
    \item{\textbf{Graph Leaner}}: For the first three parameters, we adjust these parameters respectively for u2u and i2i Graph Learners on Epinions datasets. The results are shown in Figure \ref{fig:parameters} and we can see that: (i) For u2u and i2i Graph Learners, appropriate implicit graph weight values are required. If the weight is too large, a lot of noise may be introduced leading to sub-optimal performance. Too small weight value also hurt model performance since  the learnt implicit information would become less. (ii) The increase of numbers of perspectives in GL does not necessarily lead to an increase in performance. On the contrary, too many perspectives may result in the over-fitting. (iii) As shown in \autoref{fig:search103}, the model performance reaches the best values when $L$ is 40. The performance change in the figure can indicate that too long or too short truncation  will bring loss to the model effect. The most suitable truncation length should achieve the balance between effective information and irrelevant information in the graph learning.
    \item{\textbf{Anchor-based Graph Leaner}}: For the anchor rate $\tau$, we perform experiments on a single NVIDIA Tesla V100 GPU on Ciao-28 and Epinions datasets. We record the training time (seconds) of each epoch and RMSE evaluation results. As we can see from \autoref{fig:anchor}, with the increase of the anchor rate, the performance of the model improves first and then tends to be stable, while the training time is on the rise. It can be concluded that by controlling the anchor rate within a reasonable range, the model running time can be reduced without almost loss of model performance.
\end{itemize}

\textbf{Heterogeneous Graph Neural Network}. Generally, the number $T$ of layers plays an important role for the GNN. We conduct the experiments on two datasets, and Table \ref{tab:HRGCN} presents the results of our model with different number of HGNN layers. From $T=1$ to $T=2$, the model performance is greatly improved for both datasets, which shows the necessity of the high-order interconnection. For the Epinions dataset, from $T=2$ to $T=3$, the performance still increases quickly. Generally, appropriate increase in the number of layers will make information fusion deeper. However, when $T$ is too large, the performance will drop, probably because the model introduces too much noise or becomes over-smoothing.
\begin{figure}[t]
    \centering
    \includegraphics[width=\linewidth]{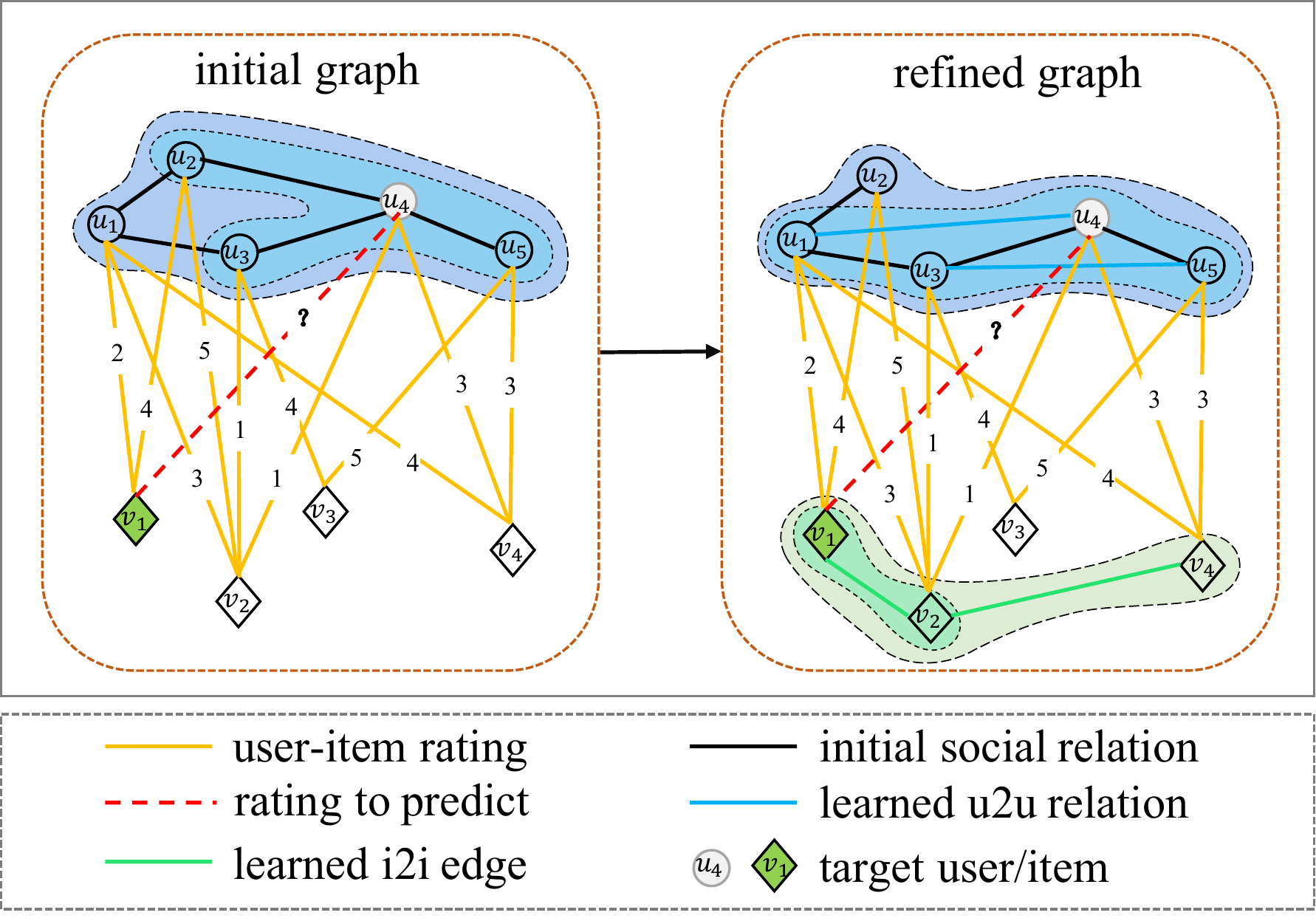}
    \caption{Visualization of an example for the case study from Ciao-5 data. Given the social connections among five users and the corresponding user-item ratings, the prediction target is the rating of user $u_4$ (white circular) on item $v_1$ (green diamond). The coverage area represents the neighboring area of the target user or item. We obtain the updated graph through the Graph Learner, based on the initial graph.}
    \label{fig:case_study}
\end{figure}
\subsection{Case Study: RQ4} 
To show the effectiveness and rationality of Global Graph Learning module, we conduct a simple case study on several users from Ciao-5 dataset. Specifically, we make a comparison between GL-HGNN and the basic GNN-based methods (HGNN) without Global Graph Learning. 

Generally, we can build a graph using user-item ratings and user social relationships as the initial graph shown in the left part of \autoref{fig:case_study}. However, we can find that there is noise in this graph structure. Although there exist social connection between user $u_4$ and $u_2$, there are huge rating differences between user $u_4$ and $u_2$ on the same item set. Besides, despite $u_4$ and $u_1$ do not have the direct social connection, their rating histories are highly overlapped. It illustrates that they may be potential friends with the similar preferences.

On the contrary,  GL-HGNN propose to adopt the Global Graph Learning  module to construct item-item connections and iteratively optimize the graph structure based on the initial graph. As shown in the the right part of \autoref{fig:case_study}, the updated graph increases the potential relationship edge and reduces noise compared with the initial graph. 
We utilize the initial graph and updated graph to make scoring predictions through HGNN, respectively. Given the ground-truth rating 2, HGNN with the updated graph (GL-HGNN) predicts the result as $2.97$, which is closer to the ground truth label compared to the value $3.44$ generated by HGNN with the initial graph. The result demonstrates the validity and rationality of our proposed Global Graph Learning module.

%% file: 3-conclusion.tex
\section{Conclusion}

In this paper, we proposed a novel method GL-HGNN to learn the  heterogeneous global graph with different relationships in a unified perspective for social recommendation. Our comparative experiments and ablation studies on four datasets illustrate that GL-HGNN can learn better graph structure with respect to social recommendation, and significantly improve the performance of recommendation. In addition, to reduce the computational complexity, we propose the Anchor-based Graph Learner.

In the future, we plan to introduce more nodes information (such as review information) for mapping multi-relation to multi-type edges in refined graph automatically.